\documentclass[12pt]{article}
\usepackage{a4wide}
\usepackage{amsmath}
\usepackage{epsfig}
\usepackage{overcite}
\makeatletter
\renewcommand\@biblabel[1]{#1)}
\def\@cite#1{\textsuperscript{#1)}} 
\makeatother
\usepackage{titlesec}
\titlelabel{\S\thetitle. \enspace}
\titleformat{\section}
{\large\sffamily\bfseries}
{\textbf{\S}\thesection.}{.5em}{}
\titleformat{\subsection}
{\large\rmfamily\itshape\mdseries}
{\thesubsection}{.5em}{}

\numberwithin{equation}{section}

\title{\sffamily\bfseries{Universal Transport Properties of Disordered Quantum Wires}}
\author{Takashi \textsc{Imamura}\footnote{E-mail:imamura@monet.phys.s.u-tokyo.ac.jp} and Miki \textsc{Wadati}
\bigskip
\\
Department of Physics, Graduate School of Science,\\
University of Tokyo,
\\
Hongo 7-3-1, Bunkyo, Tokyo 113-0033, Japan}

\begin{document}
\large
\setlength{\baselineskip}{21pt}
\maketitle
\begin{abstract}
For disordered quantum wires
which belong to all ten universality classes,   
the universal quantities of transport properties
are obtained through DMPK approach.
Calculated are the  universal parts of one- and two-point correlation
functions for probability distribution functions
of transmission eigenvalues. In this analysis,
the asymptotic solution of DMPK equation is used.
Transport properties for
new universality classes(chiral and Bogoliubov-de Gennes
classes) are discussed comparing with those for standard class.
\end{abstract}

\bigskip

\section{Introduction}
In disordered and mesoscopic systems, one of the most important phenomena is
the universal behavior which is determined by the symmetries of the 
Hamiltonian and by the dimensionality of the system. 
For instance the conductance 
fluctuation in mesoscopic system is known to
be independent of the conductance itself and the shape of the conductor,
while it depends on whether or not the magnetic field are added in
the system.~\cite{Alt ,LeSt} 
Another example is a localization property of disordered
system. The scaling theory of Anderson localization indicates that
the transition only occur at three dimension regardless of a local Hamiltonian of the
system.~\cite{AALR}
 
The Hamiltonians of disordered system can be classified to ten types by 
time-reversal symmetry, spin-rotation symmetry and so on
which are not spatial. This classification follows from 
one-to-one correspondence of these Hamiltonians to Riemannian symmetric 
spaces in mathematical terminology.~\cite{Zir}
For the correspondence to actual physical system,
these ten classes can be divided into three classes, 
standard, chiral and Bogoliubov-de 
Gennes (B-dG)classes. 
The standard class corresponds to the system which has a 
disordered potential. The chiral class is applied to the Dirac particles
in a random gauge field and the tight binding model with
random nearest-neighbor hopping term. The B-dG class is found out 
when the system obeys the Bogoliubov-de Gennes equation.~\cite{AlZi}
Two dimensional
random bond Ising model is also classified to this 
class.~\cite {GRL,BSZ}

\begin{figure}[htp]
\begin{center}
\epsfig{file=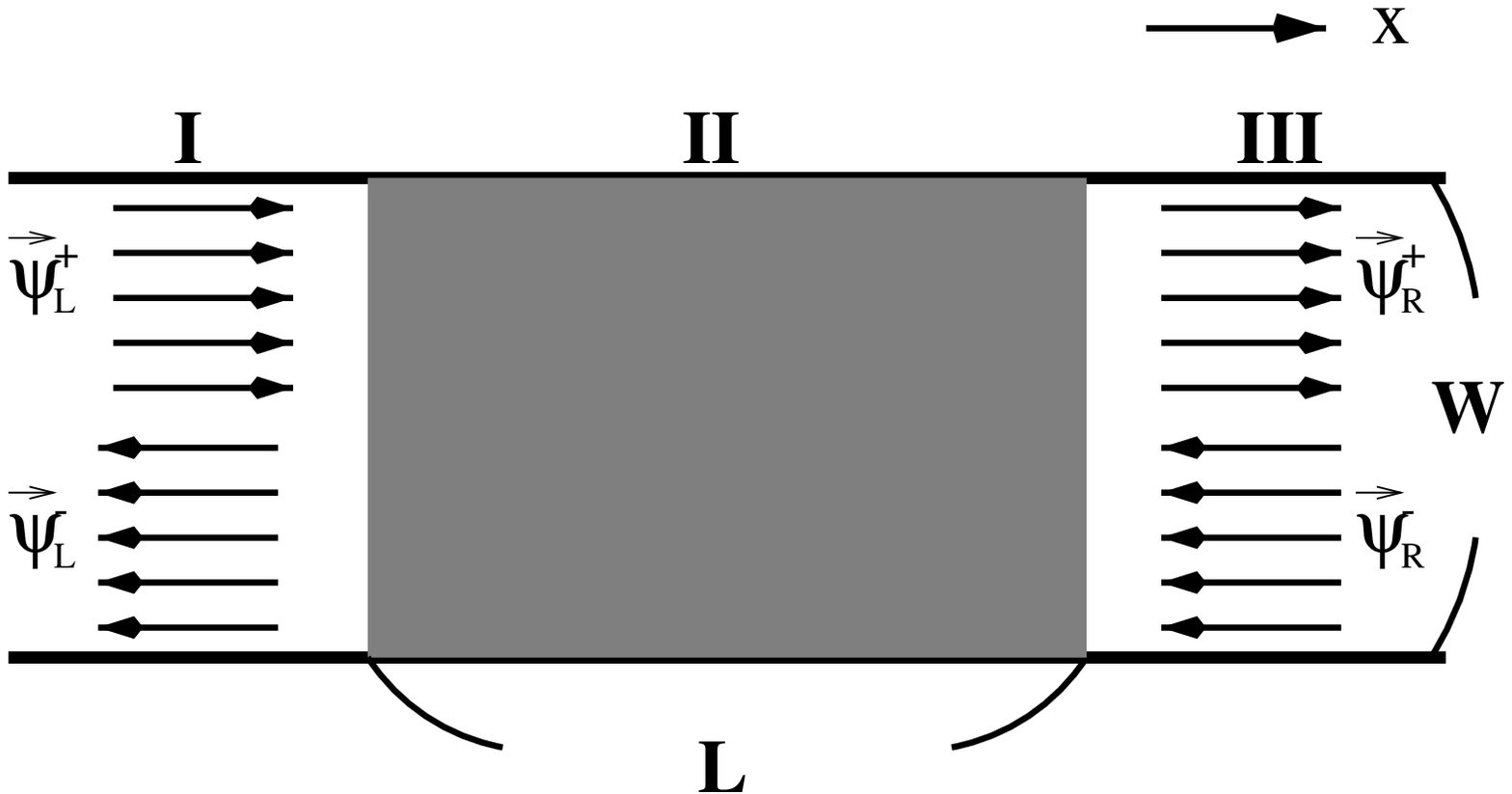}
\end{center}
\caption{The disordered quantum wire. \label{fig:c}The wire
can be divided into three regions I,II and III. I and
III represent the normal leads and II is disordered region. 
There are several propagating modes, which are represented by arrows,
depending on the transverse length W.}
\begin{center}
\epsfig{file=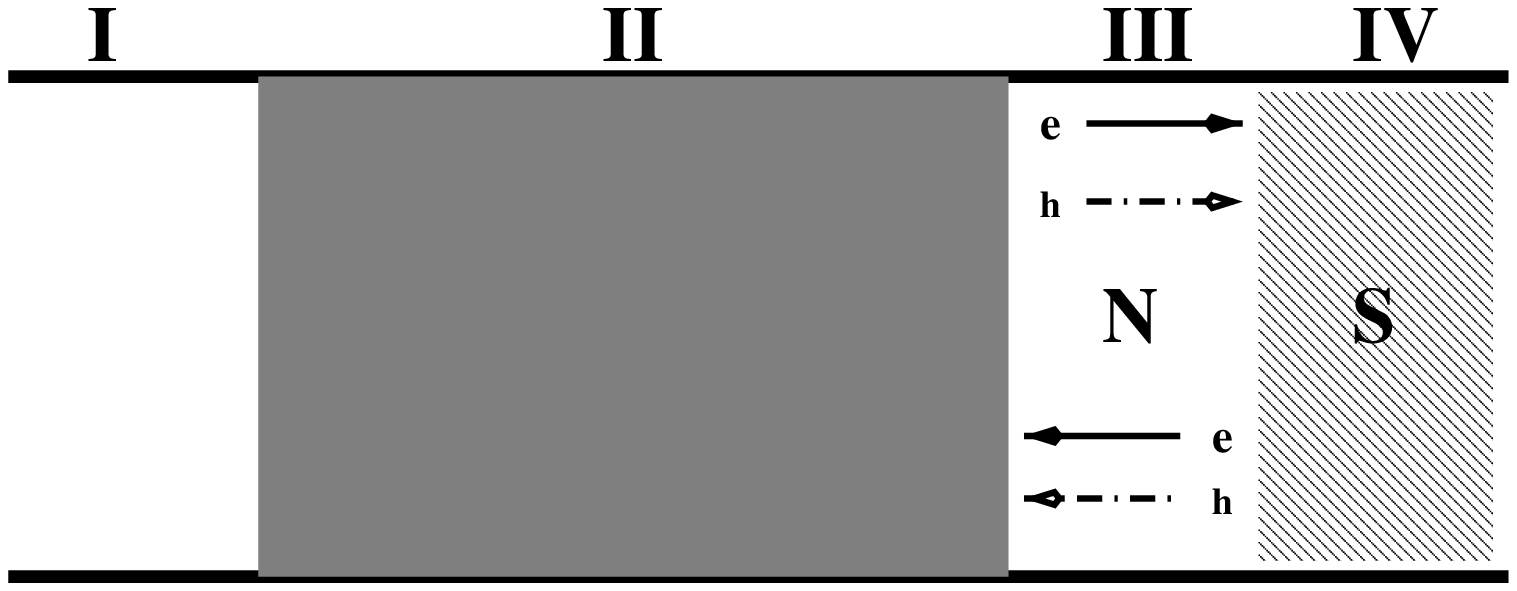}
\end{center}
\caption{The disordered quantum wire containing a normal-superconducting
junction. \label{fig:d} The Andreev reflection, where 
the electron(e) is reflected as a hole(h) with the opposite velocity,
occurs at the interface between the region III(a normal lead) 
and IV(superconducting region). 
The conductance can be expressed as eq.~\eqref{eq:Gns}
for the case where the time-reversal symmetry is satisfied.}
\end{figure}

This classification is believed to be correct regardless of the dimensionality.
We, however, concentrate on the quantum wire (quasi-one dimensional 
system). The quantum wire we consider is illustrated in Fig.~\ref{fig:c}. 
Different
from one dimensional system, there are several propagating modes 
$\psi_n$,  
\begin{gather}
\psi_{n}^{\pm} = c_{n}\phi_{n}(y,z)\exp(\pm ik_nx), 
(1\le n\le Nd),
\end{gather}
where $N$ is the number of different propagating 
states and $d$ is the number of degeneracies.
Another feature of a quantum wire is that transport properties change 
whether or not $L$ in Fig.~\ref{fig:c} is longer than the localization length,
$\xi$. If
$L\gg\xi$ , it behaves like an insulator, while
it behaves like a metal if $L\ll\xi$.

For mesoscopic systems, the transport quantities can be calculated by use
of the transfer matrix as follows.
The transfer matrix $\cal M$ that connects left side states 
with right side states is 
represented as
\begin{gather}
\left(\begin{array}{c}
         \vec{c^{+}_{\rm{R}}} \\
         \vec{c^{-}_{\rm{R}}}
        \end{array}\right)
         = \cal {M}
         \left(\begin{array}{c}
         \vec{c^{+}_{\rm{L}}} \\
         \vec{c^{-}_{\rm{L}}}
        \end{array}\right). 
\end{gather}
Due to 
the flux conservation of the system,
the matrix $\cal M$ belongs to pseudo-unitary group ${\cal U}(N,N) $. 
Provided that the system satisfies additional symmetries 
(time reversal, spin rotation, chiral etc.), $\cal M$
belongs to the subgroup of ${\cal U}(N,N) $. These subgroups
to which each universality classes belongs and the symmetries satisfied
are summarized in Table~\ref{tab:a}. 
The transfer matrices can be decomposed to 
relevant and irrelevant degrees of freedom,
\begin{gather}
\label{eq:Trans}
{\cal M_{\rm L}} = 
\left(\begin{array}{cc}
         {\cal A} & 0 \\
         0 & {\cal B}
        \end{array}\right)
         \left(\begin{array}{cc}
         \cosh \rm X & \sinh \rm X \\
         \sinh \rm X & \cosh \rm X
         \end{array}\right) 
\left(\begin{array}{cc}
         {\cal C}& 0 \\
         0 & {\cal D}
        \end{array}\right),
\end{gather}
where ${\cal A} \sim {\cal D}$ are $N \times N$ unitary matrices and 
X is an $N \times N$ diagonal matrix with eigenvalues $x_j(1\le j\le N)$.
The eigenvalues $x_j$ are relevant variables while matrix elements of 
 ${\cal A} \sim {\cal D}$ are irrelevant ones, because
transport quantities are expressed in terms of only $x_j$. In this paper, we
consider the following three quantities.
\begin{enumerate}
\item conductance $G$ :
\begin{gather}
  \label{eq:G}
  G=d\sum_{n=1}^N \frac{1}{\cosh ^2 x_{n}}, 
\end{gather}
\item shot-noise power $P_{\text{shot}}$ :
\begin{gather}
 \label{eq:P}
  P_{\text{shot}}=d\sum_{n=1}^N \frac{1}{\cosh ^2 x_{n}}
(1-\frac{1}{\cosh ^2 x_{n}}),  
\end{gather}
\item conductance $G_{\text{NS}}$ for a system with a 
normal-superconducting junction :
\begin{gather}
  \label{eq:Gns}
  G_{\text{NS}}=d\sum_{n=1}^N \frac{1}{\cosh ^2 2x_{n}}.
\end{gather}
\end{enumerate}  
Equation~\eqref{eq:G} is called the multi-channel Landauer formula
and was derived in this form in ref.~\citen{FiLe}. The shot-noise 
power means the time fluctuation of the current and eq.~\eqref{eq:P}
was first obtained in ref.~\citen{But}. When the system has a  normal 
metal-superconducting junction, the reflection of particles
on this interface
becomes anomalous (Andreev reflection), and the Landauer formula
is changed to eq.~\eqref{eq:Gns}.~\cite{Bee1} See Fig.~\ref{fig:d}.
Note that this formula is applicable only if 
the time-reversal symmetry is satisfied.
In the case where the time-reversal symmetry is broken,
eq.~\eqref{eq:Gns} is not correct and we do not consider such case
in this paper.

The transport quantities we are interested in are statistical quantities 
which can be written in terms of correlation functions obtained from 
the probability distribution  $P(x)\equiv P(x_1,\cdots x_N)$ over 
impurity configurations. In general one represents  
a transport quantity $A$ as
\begin{gather}
\label{eq:TQ}
  A\equiv
  d \sum_{n=1}^N f(x_n).
\end{gather}
Then the average $\langle A\rangle$ and the variance $\rm Var A$ for 
impurity configurations are given by  
\begin{align}
  \label{define_var_A}
\langle A\rangle &= d\int
  \mathrm{d} x f(x)K_1(x), \\
  \text{Var}A
  &=\left<A^2 \right>-\left<A\right>^2 \nonumber \\
  &=
  d^2 \, \int\int
  \mathrm{d} x  \mathrm{d}x^{\prime}f(x)
         f(x')K_2(x,x') ,
\end{align}
where $K_1$ and $K_2$ are one- and two-point correlation functions respectively,
\begin{align}
  K_1(x_1)
 &=N \int\cdots
  \int
  P(x) \,
  \mathrm{d}x_2  \, \mathrm{d}x_3
  \cdots \mathrm{d}x_N, \\
  K_2(x_1,x_2)
  &=N(N-1)\int\cdots
  \int
  P(x) \,
  \mathrm{d}x_3  \, \mathrm{d}x_4
  \cdots \mathrm{d}x_N.
\end{align}
We expand  $K_1$ and $K_2$ by
$\frac{1}{N}$.
The term of $(\frac{1}{N})^0$ represents the universal 
behavior. In $K_2$, we see that the first term in the expression
is of the order
$(\frac{1}{N})^0$. Thus the fluctuation of the transport quantity
is universal. On the other hand, in $K_1$ the first term is 
$(\frac{1}{N})^{-1}$ and the second is $(\frac{1}{N})^0$.  
The first term can be interpreted as the average in which propagating 
modes are regarded as classical particles, while the second 
exhibits the quantum
correction.

We shall obtain the universal terms of $K_1$ and $K_2$ for all ten universality
classes with respect to a quantum wire. Our main results are 
eq.\eqref{eq:A}, eq.\eqref{eq:B}, eq.\eqref{eq:C} and eq.\eqref{eq:D}.  
Those universal terms 
have already been obtained for standard class but not for chiral and B-dG
classes. The purpose of this paper is to clarify how the difference of 
universality classes influences the universal behavior of transport
properties.

The form of $P(x)$ depends on the dimensionality of the system. 
In a quantum dot
(zero dimensional system) 
this becomes the ensemble of the random matrix theory(RMT),~\cite{Imry,Me}
\begin{eqnarray}
  P(x)=\prod_{i<j}^{}|x_i-x_j|^{\beta} \cdot
  \exp\left(-\sum_{i=1}^N
                W(x_i)\right) .
\end{eqnarray}
In RMT, $n$ -point correlation functions ($n\ge 2$) are independent of $W(x)$
in case that $N$ becomes infinity. And they can be evaluated exactly.~\cite{BrZe,Be,Fo} 
In a quantum wire, the probability distribution is known to satisfy the following 
scaling equation called DMPK equation.
This was first obtained in standard classes.~\cite{Dr, MPK, MeSt, MaCh}
Recently these equations for chiral and B-dG classes were obtained,~\cite{MBF1,MBF2,BFGM} 
\begin{gather}
  \label{eq:1}
  \frac{\partial P(x;s)}{\partial s}=\frac{1}{2\gamma}\sum_{j=1}^N \frac{\partial}
  {\partial x_j}J\frac{\partial}{\partial x_j}J^{-1}P(x;s),~~s=\frac{L}{\ell},
\end{gather}
where $\ell$ is the mean free path, and 
$J$ and $\gamma$ depend on universality classes :
\begin{itemize}
\item Standard and B-dG classes
\begin{gather}
  J=\prod_{j=1}^N \sinh^\kappa (2x_j)\prod_{j>k}{}\sinh^\beta
  (x_j - x_k)\sinh^\beta(x_j + x_k), \\
   \gamma = \beta N +1+\kappa -\beta, \\
   0\le x_j\le \infty.
\end{gather}
\item chiral class
\begin{gather}
  J=\prod_{j>k}{}\sinh^\beta
  (x_j - x_k), \\
   \gamma = \frac{1}{2}\left (\beta (N -1)+2\right ),\\
   -\infty \le x_j \le \infty.
\end{gather} 
\end{itemize}
The value of $\beta$ and $\kappa$ are listed in Table~\ref{tab:a}.
Note that the domains of $x_j$ are different for chiral and 
other classes. This is because in chiral class the value 
of $x_j$ is determined when the transfer matrix is fixed,
but in other classes the sign of $x_j$ can not be determined.
The sign can be absorbed in the irrelevant part 
( ${\cal A} \sim {\cal D}$ in eq.~\eqref{eq:Trans})
in standard and 
B-dG classes fixing the transfer matrix.

We like to make a remark on eq.~\eqref{eq:1}.
In ref.~\citen{Brouwer}, Brouwer et.al. pointed out that 
for general chiral class, DMPK equation depends on another parameter
in addition to 
the mean free path. The additional parameter
follows from the geometric structure of the transfer matrices.

The DMPK equation has a remarkable structure in mathematical point of view.
It can be mapped into imaginary time Schr\"{o}dinger equation with the
Calogero-Sutherland Hamiltonian and is related to the diffusion 
in the corresponding symmetric space that is the coset between the group
to which the transfer matrix belongs and its maximal compact 
subgroup.~\cite{Hu,Ca}
 These cosets
are described in Table~\ref{tab:a}.~\cite{Hel}

\begin{table}[htbp]
  \begin{center}
 \begin{tabular}{|c|c|c|c|c|c|c|c||c|c|c|}         \hline
      class & $H$ & TR & SR & $\beta$ &$\kappa$ & $d$ &$N$&$\cal M$& $\cal L$ & $\cal M/\cal L$
      \\ \hline \hline
      & AI & $\bigcirc$ & $\bigcirc$ &1&1&2&$M$&$\mathit{Sp}(N,R)$&$U(N)$&CI\\ \cline{2-11}
      standard  & A  & $\times$    &$\bigcirc\left(\times\right)$ &2&1&2(1)&$M(2M)$&$\mathit{U(N,N)}$
&$\mathit{U(N)\times U(N)}$
&AIII   \\ \cline{2-11}
      & AII & $\bigcirc$ & $\times$ &4&1&2&$M$&$\mathit{SO^\ast(4N)}$&$U(2N)$&DIII     \\ \hline
      & CI & $\bigcirc$ & $\bigcirc$ &2&2&4&$M$&$\mathit{Sp(N,c)}$& $\mathit{Sp(N)}$&C      \\ \cline{2-11}
      BdG    & C    & $\times$    &$\bigcirc$ &4&3&4&$M$&$\mathit{Sp(N,N)}$&$\mathit{Sp(N)\times Sp(N)}$&CII     \\ \cline{2-11}
      & DIII    & $\bigcirc$ &$\times$ &2&0&2&$2M$&$O(2N,C)$&$O(2N)$&D       \\ \cline{2-11}
      & D   & $\times$ &$\times$ &1&0&1&$4M$&$O(N,N)$&$O(N)\times O(N)$&BDI       \\ \hline
      & BDI  & $\bigcirc$ &$\bigcirc$ &1&-&2&$M$&$\mathit{GL(N,R)}$&$O(N)$&AI       \\ \cline{2-11}
   chiral   & AIII   & $\times$ &$\bigcirc(\times)$ &2&-&2(1)&$M(2M)$&$\mathit{GL(N,C)}$&$U(N)$&A       \\ \cline{2-11}
      & CII   & $\bigcirc$ &$\times$ &4&-&2&$M$&$U^\ast (N)$&$\mathit{Sp(N)}$&AII       \\ \hline
    \end{tabular}

    \caption{Classification of the DMPK equation.
      In the second column $H$  is the Cartan symmetric class, which
      denotes
      the symmetry class of
      the Hamiltonian of 
      the quantum wire.
      Every
      class is classified by the  time-reversal symmetry (TR) and spin-rotational symmetry  (SR):
      $\bigcirc$ and $\times$ denote
      whether
      the system preserves the symmetry or not.
      Parameters
      $\beta$ and $\kappa$ characterize
      each universality class in $J$ in eq.~(\ref{eq:1}).
      $d$ is the number of degeneracy of the transmission 
      eigenvalue.
      The column labeled by $N$ represents the number of distinct propagating modes.
      $M$ in the column depends on the width W of the system illustrated in Figure~\ref{fig:c}.
      Note that the total number of propagating modes $Nd$ is the same for each universality 
      class though $N$ varies with each class.  
      The transfer matrices $\cal M$ belong to the pseudo-unitary group $U(N,N)$ because of 
      the flux conservation. Provided the system preserves additional symmetries (time-reversal,
      spin-rotation, chiral etc.), they reduce to the subgroup of $U(N,N)$. 
      $\mathcal L$ denotes the maximal compact subgroup of each transfer matrix.
       The DMPK equation is related to the diffusion on
      the symmetric space $\cal M/\cal L$. In this table, we follow the notations of Lie group in ref~\citen{Hel}.}
    \label{tab:a} 
  \end{center}
\end{table}

This paper is arranged as follows.
In section 2, we examine the asymptotic
solutions of DMPK equation in metallic regime.
From these asymptotic solutions
we calculate the universal
parts of one- and two-point correlation functions in section 3.
For detail analysis we employ the asymptotic formula
by Dyson and Beenakker and the method of functional derivative.
In section 4, we discuss the results for chiral and B-dG classes 
comparing with those for standard class. 
    
\section{The Asymptotic Solution of DMPK Equation}
We investigate the asymptotic solutions of DMPK equations for metallic 
regime.
The initial condition of the solutions for $s=0$ is 
\begin{gather}
\label{initial}
P(x;s=0) =\prod_{i}^{N}\delta(x_i),
\end{gather}
due to the fact that when $s=0$, a conduction is absolutely 
ballistic.
As we have explained above,
in the metallic regime 
the size of the system, $L$, is much longer than the 
localization length $\xi$. In a quasi-one dimensional system,
$\xi$ can be estimated to be an order $N\ell$. 
Thus the condition for metallic regime is 
$s\ll N$.
We generalize the method of ref.\citen{Ca} 
to the B-dG and Chiral 
classes.

To reveal  the asymptotic behavior, 
it is useful to define the   ``wave function'' $\Psi$ as
 \begin{gather}
   P(x;s)=J^{1/2}\Psi(x;s) .
 \end{gather}
Then we can map the DMPK equation~(\ref{eq:1}) onto the (imaginary
time) Schr\"{o}dinger equation,
\begin{equation}
  \label{eq:Sch}
  -2\gamma\frac{\partial \Psi(x;s)}{\partial s}=
  \mathcal{H} \, \Psi(x;s).
\end{equation}
The explicit form of the ``Hamiltonian'' is given as follows. 
\begin{itemize}
\item Standard and B-dG classes
\begin{multline}
  \label{eq:HsB}
  \mathcal{H}\equiv
  -\sum_{j=1}^N\frac{\partial^2}{\partial x^2_j}+{\kappa(\kappa-2)}\sum_
  {j=1}^N\frac{1}{\sinh^22x_j}
  \\
  +\frac{\beta(\beta-2)}{2}\sum_{j<k}^{}\left.(\frac{1}{\sinh^2(x_j-x_k)}
    +\frac{1}{\sinh^2(x_j+x_k)}\right.)+\text{const.}
\end{multline}
\item chiral class
\begin{gather}
  \label{eq:Hc}
  \mathcal{H}\equiv
  -\sum_{j=1}^N\frac{\partial^2}{\partial x^2_j}+
    \frac{\beta(\beta-2)}{2}\sum_{j<k}^{}\frac{1}{\sinh^2(x_j-x_k)}
    +\text{const.}
\end{gather}
\end{itemize}
These Hamiltonians, \eqref{eq:HsB} and \eqref{eq:Hc}, are called Calogero-Sutherland model (C-S 
model) of type $C_N$
and $A_N$ respectively and are known to be integrable.~\cite{DiVi,OlPe,WNUK} The C-S model
is related to the radial part of the Laplace-Beltrami Operator 
on the symmetric space, $\cal O_{\cal B}$;~\cite{OlPe}
\begin{gather}
  \label{eq:H}
  \mathcal{H} = 
  J^\frac{1}{2}{\cal O_{\cal B}}J^\frac{-1}{2}.
\end{gather}
Thus DMPK equation can be expressed in terms of $\cal O_{\cal B}$ as
\begin{gather}
  \label{eq:Bdif}
  \frac{\partial P(x;s)}{\partial s}=\frac{1}{2\gamma}J{\cal O_{\cal B}}J^{-1}P(x;s) .
\end{gather}
The eigenfunction  $\Phi _k$ of $\cal O_{\cal B}$ with eigenvalue $k^2$ 
is known in mathematics as
zonal spherical function~\cite{Ma}. Then,  
the general solution of eq.~\eqref{eq:Bdif} for
ballistic initial condition~\eqref{initial} is 
\begin{gather}
P(x;s)= J\sum\exp(-\frac{sk^2}{2\gamma})\Phi _k.
\end{gather}
Here the summation is taken 
for all eigenstates. This formula can be regarded as the Fourier
transform on symmetric spaces and can be changed
as follows,~\cite{Har, HeOp}
\begin{equation}
  \label{general_wf}
  P(x;s)
  =
  J\iint \prod_i \mathrm{d} k_i \,
  \mathrm{e}^{- \frac{s}{2 \gamma} \sum_i k_i^{2}} \, \Phi_k(x)
  \frac{1}{|c(k)|^2} .
\end{equation}
The Harish-Chandra  $c$-function, $c(k)\equiv c(k_1,\cdots k_N)$, is given by
\begin{equation*}
  c(k)
  =
\begin{cases}
\displaystyle
  \prod_{i<j}
  \frac{\Gamma(\mathrm{i} \frac{k_i - k_j}{2}) \,
      \Gamma(\mathrm{i} \frac{k_i+k_j}{2})}{
      \Gamma(\frac{\beta}{2} + \mathrm{i} \frac{k_i -k_j}{2}) \,
      \Gamma(\frac{\beta}{2} + \mathrm{i} \frac{k_i +k_j}{2})
      }
  \cdot
  \prod_j
    \frac{\Gamma(\mathrm{i} \frac{k_j}{2})}{
      \Gamma(\frac{\kappa}{2} + \mathrm{i} \frac{k_j}{2})
      } & \text{for standard and B-dG},
 \\[2mm]
\displaystyle
  \prod_{i<j}
  \frac{\Gamma(\mathrm{i} \frac{k_i - k_j}{2})}{
      \Gamma(\frac{\beta}{2} + \mathrm{i} \frac{k_i -k_j}{2})}& \text{for chiral}.
\end{cases}
\end{equation*}

In the metallic regime, we can assume $|x|\gg 0$ and the momentum $|k|\gg 0$.
Applying these two conditions, we have 
\begin{gather}
\label{eq:azona}
\Phi_k(x)\sim J^{-\frac{1}{2}}\sum_{r\in W}c(rk)\exp (i(rk,x))
\end{gather}
where $W$ is Coexter group of the corresponding root system and $(,)$
is the inner product, and the asymptotic form for $\Gamma$-function is used ;
\begin{gather}
\label{eq:agamma}
\frac{\Gamma(\frac{\beta}{2}+iy)}{\Gamma(iy)} \sim |y|^{\frac{\beta}{2}}
\exp(\frac{i\pi\beta}{4}) ~~~~{\text{for}}~~\beta\in \{1,2,4\},y\rightharpoonup
\infty.
\end{gather}
Using eq.~\eqref{eq:azona} and eq.~\eqref{eq:agamma} in eq.~\eqref{general_wf},
we arrive at the following expressions.  
\begin{itemize}
\item Standard and B-dG classes
\begin{gather}
\label{eq:AsB}
  P(x;s)
  \simeq
  \prod_{i<j}^{}|\sinh^2x_j-\sinh^2x_i|^{\frac{\beta}{2}}
  |x^2_j-x^2_i|^{\frac{\beta}{2}}\prod_{i=1}^N\exp(-\frac{x^2_i\gamma}{2s})
  \bigl(
  x_i \sinh (2x_i)
  \bigr)^{\frac{\kappa}{2}} ,
\end{gather}
\item Chiral class
\begin{gather}
\label{eq:Ach}
  P(x;s)
  \simeq
  \prod_{i<j}^{}(x_j-x_i)^{\frac{\beta}{2}}\sinh^{\frac{\beta}{2}}(x_j-x_i)
  \prod_{i=1}^N\exp(-\frac{x^2_i\gamma}{2s})
   .
\end{gather}
\end{itemize}
In both cases the interaction term of $P(x)$ differs from the  
one for RMT which has a form of $|x_i-x_j|^{\beta}$.
It is this difference that changes the transport properties between
quantum dot and quantum wire. 

\section{Correlation Function}
\subsection{Asymptotic Formula }
For calculating universal terms in $K_1$ and $K_2$, we use the 
asymptotic formula given by F.J.Dyson.~\cite{Dys}

For the case that the probability distribution
$P(x)$ is the ensembles of RMT,
\begin{align}
P(x) &= \exp(-\beta W(x)),\\
   W(x)  &= -\sum_{i<j}{\rm ln}|x_j - x_i| +\sum_{i}V(x),
\end{align}
Dyson derived the following integral equation,
\begin{gather}
\label{eq:Asy}
\int {\rm d}x'K_1(x'){\rm ln}|x-x'|
+\frac{1}{2}(1-\frac{2}{\beta}){\rm ln}K_1(x)
=
V(x) + {\rm const.}
\end{gather}
It is generalized easily 
for $W(x)$ having a non-logarithmic repulsion term 
$\delta u$\cite{Bee2},
\begin{align}
W(x)  &= -\sum_{i<j}u(x_j,x_i) +\sum_{i}V(x) ,\\
u(x,x')&= {\rm ln}|x-x'| +\delta u.
\end{align}
The generalization of eq.~\eqref{eq:Asy} to the non-logarithmic repulsion is
\begin{gather}
\label{eq:Iasy}
\int {\rm d}x'K_1(x')u(x,x')
+\frac{1}{2}(1-\frac{2}{\beta}){\rm ln}K_1(x) +\frac{1}{2}\delta u(x,x)
=
V(x) + {\rm const.},
\end{gather}
where we assume that 
$\delta u(x,x')$ has a limit value 
$\delta u(x,x')\bigl|_{x'\rightharpoonup x}$.

\subsection{One-Point Correlation Function $K_1$}
\subsubsection{Standard and B-dG Classes}
In the case of standard and B-dG
classes,
$u(x,x')$, $\delta u(x,x')$ and $V(x)$  
are due to eq.~\eqref{eq:AsB},
\begin{align}
\label{eq:sBu}
u(x,x') &= {-\frac{1}{2}}{\rm ln}[(x-x')\sinh (x-x')]{-\frac{1}{2}}
{\rm ln}[(x+x')\sinh (x+x')], \\
\delta u(x,x') &= {-\frac{1}{2}}{\rm ln}[(x-x')^{-1}
\sinh (x-x'){-\frac{1}{2}}]
{\rm ln}[(x+x')\sinh (x+x')], \\
\delta u(x,x) &= {-\frac{1}{2}}{\rm ln}[(2x)\sinh (2x)],\\
V(x) &= \frac{1}{2s}(N-1+\frac{1+\kappa}{\beta})x^2 -\frac{\kappa}{2\beta}
{\rm ln}|2x\sinh 2x|.
\end{align} 
Substituting them into eq.~\eqref{eq:Iasy}, we find
\begin{gather}
\label{sBa}
-\int_{0}^{\infty}{\rm d}x' K_1(x')
({-\frac{1}{2}}{\rm ln}(x-x')\sinh (x-x'){-\frac{1}{2}}
{\rm ln}(x+x')\sinh (x+x') ) \nonumber\\ 
+\frac{1}{2}(1-\frac{2}{\beta}){\rm ln}K_1 
= \frac{1}{2s}(N-1+\frac{1+\kappa}{\beta})x^2
+\frac{\beta-2\kappa}{2\beta}{\rm ln}|2x\sinh 2x|+{\text{const.}}.
\end{gather}
We expand $K_1(x)$ by $N$,
\begin{eqnarray}
K_1= K_{1,N} + K_{1,0} + \cdots~~,  
\end{eqnarray}
where $K_{1,N}$ is a function of the order $N$ and $K_{1,0}$ is of the order $N^0$.
The equation satisfied by $K_{1,N}$ is due to eq.~\eqref{sBa},
\begin{gather}
-\int_0^{\infty}{\rm d}x' K_{1,N}(x')({-\frac{1}{2}}
{\rm ln}(x-x')\sinh (x-x'){-\frac{1}{2}}
{\rm ln}(x+x')\sinh (x+x') ) \nonumber\\ 
= \frac{N}{2s}x^2 + {\rm const.}~~.
\end{gather}
The solution of this integral equation in metallic regime $s\gg 1,s\gg x$
is~\cite{MePi}
\begin{gather}
\label{eq:ohm}
K_{1,N}(x)= \frac{N}{s}\Theta (s-x) =\frac{NL}{l}\Theta (s-x),
\end{gather}
where $\Theta (x)$ is the step function defined by
\begin{gather}
\Theta (x) = 
\begin{cases}
\displaystyle 
1
&\text {for}~~  x\ge 0 \\
\displaystyle 
0
&\text {for}~~  x < 0.
\end{cases}
\end{gather}
Since $K_{1,N}$ is proportional to the length $L$ of the system
and the number $N$ of the modes, we may interpret it as the 
one-point function where the modes are regarded to be the classical
particles. Averaging the conductance~\eqref{eq:G} with respect to
eq.~\eqref{eq:ohm}, we see that it agrees with the Ohm's law. 

The equation satisfied by $K_{1,0}$ is
\begin{gather}
\label{eq:sBK1}
\int_{-\infty}^{\infty}{\rm d}x' K_{1,0}(x'){\rm ln}|(x-x')\sinh(x-x')|=
\frac{\beta - 2\kappa}{2\beta}{\rm ln}|2x\sinh 2x|.  
\end{gather}
Here, the term $\frac{x^2}{s}$ is neglected because $s\gg x$, and
the term ${\rm ln}K_{1,N}$ is absorbed to the constant term. We change
the range of integration requiring $K_{1,0}(x) = K_{1,0}(-x)$. This 
integral equation can be solved readily by the Fourier expansion. The solution
of eq.~\eqref{eq:sBK1} is
\begin{gather}
\label{eq:A}
K_{1,0}(x) =\frac{\beta - 2 \kappa}{\beta}
  \left(
    \frac{1}{4} \delta(x-0^+) +
    \frac{1}{4 x^2 + \pi^2}
  \right)
\end{gather}
\subsubsection{Chiral Class}
According to eq.~\eqref{eq:Ach}, $u(x,x'),\delta u(x,x')$ and $V(x)$ are respectively given by
\begin{align}
\label{eq:chu(x,x)}
u(x,x') &=-\frac{1}{2}{\rm ln}|(x-x')\sinh (x-x')|,\\
\delta u(x,x') &=-\frac{1}{2}{\rm ln}|(x-x')^{-1}\sinh (x-x')|,\\
\delta u(x,x) &= 0\\
V(x) &= \frac{1}{4s}(N-1+\frac{2}{\beta})x^2.
\end{align}
Substituting them into eq.~\eqref{eq:Iasy}, we find
\begin{gather}
{\frac{1}{2}}\int_{-\infty}^{\infty}{\rm d}x' K_1(x')
{\rm ln}(x-x')\sinh (x-x')
= \frac{1}{4s}(N-1+\frac{2}{\beta})x^2 +{\text{const.}}
.
\end{gather}
Expanding $K_1$ by $N$, we have the equation for $K_{1,N}$,
\begin{gather}
{\frac{1}{2}}\int_{-\infty}^{\infty}{\rm d}x' K_{1,N}(x')
{\rm ln}(x-x')\sinh (x-x')
= \frac{N}{4s}x^2 + {\rm const.}~~.
\end{gather}
The solution of $K_{1,N}$ in the metallic regime is given by
\begin{gather}
K_{1,N}(x)= \frac{N}{2s}\Theta (s-|x|).
\end{gather}
The equation for $K_{1,0}$ is
\begin{gather}
\int_{-\infty}^{\infty}{\rm d}x' K_{1,0}(x')
{\rm ln}|(x-x')\sinh (x-x')|
=0,
\end{gather}
and simply the solution is 
\begin{gather}
\label{eq:B}
K_{1,0}(x') = 0.
\end{gather} 
\subsection{Two-Point Correlation Function $K_2$}

\subsubsection{Standard and B-dG Classes}
The two-point correlation function $K_2$ is related to the one-point function
$K_1$ through the functional derivative,
\begin{gather}
\label{eq:Fde}
K_2(x,x')=\frac{1}{\beta}\frac{\delta K_1(x)}{\delta V(x')},
\end{gather}
or, the integral equation,
\begin{gather}
\label{eq:K2}
\delta K_1(x)=\int_{0}^{\infty}{\rm d}x'\beta K_2(x,x')\delta V(x').
\end{gather} 
Due to eq.~\eqref{eq:Iasy},
the integral equation for $\delta K_{1,N}(x)$ is 
\begin{gather}
\label{eq:cor}
-\int_{0}^{\infty}{\rm d} x'\delta K_{1,N}(x')u(x,x')=\delta V(x) +\rm{const.}~~.
\end{gather}
Here the constant term should be zero so as to fulfill the conservation of 
the probability,
\begin{gather}
\int_{0}^{\infty}{\rm{d}}x\delta K_{1,N}(x) =0.
\end{gather}
Substituting eq.~\eqref{eq:sBu} into eq.~\eqref{eq:cor}, we have
\begin{gather}
\label{eq:sBF}
\int_{-\infty}^{\infty}{\rm{d}} x'\delta K_{1,N}(x') 
{\cal U}(x-x') =2\delta V(x),
\end{gather}
where
\begin{gather}
\label{eq:calU}
{\cal{U}}(x) = {\rm{ln}}|2x\sinh (x)|.
\end{gather}
This integral equation can be solved by convolution. The result is
\begin{align}
\label{eq:sBconb}
\delta K_{1,N} &= \int_{0}^{\infty}{\rm d}x' \beta 
\left({\cal K}(x-x')+{\cal K}(x+x')\right)\delta V(x'),\\
{\cal K}(x)&= \frac{2}{\beta\pi}\int_{0}^{\infty}\frac{\cos kx}{{\cal U}(k)},
\label{eq:calK}
\end{align}
where ${\cal U}(k)$ is the Fourier transformation of  ${\cal U}(x)$,
\begin{gather}
{\cal U}(k) = -\frac{\pi}{|k|}
\left(1+\coth \left(\frac{1}{2}\pi |k|\right)\right).
\end{gather}
Thus, from eq.~\eqref{eq:K2} and eq.~\eqref{eq:sBconb},
the two-point correlation function for standard and B-dG classes
is 
\begin{gather}
\label{eq:C}
K_2(x,x') ={\cal K}(x-x')+{\cal K}(x+x').
\end{gather}

\subsubsection{Chiral Class}
In chiral classes, the equations corresponding to eq.~\eqref{eq:K2}
 and eq.~\eqref{eq:cor} are 
respectively
\begin{align}
&\delta K_1(x)=\int_{-\infty}^{\infty}{\rm d}x'K_2(x,x')\delta V(x'),\\
\label{eq:ccor}
&-\int_{-\infty}^{\infty}{\rm d}x'\delta K_{1,N}(x')u(x,x') = \delta V(x)
+{\rm const.}~~.
\end{align}
Substitution of eq.~\eqref{eq:chu(x,x)} into eq.~\eqref{eq:ccor} gives
\begin{gather}
\int_{-\infty}^{\infty}{\rm{d}} x'\delta K_{1,N}(x') 
{\cal U}(x-x') =2\delta V(x)+{\rm const.}~~.
\end{gather}
This integral equation is the same as eq.~\eqref{eq:sBF} for standard and B-dG
classes. Therefore, the two-point function $K_2$ for chiral class is due to 
eq.~\eqref{eq:K2},
\begin{gather}
\label{eq:D}
K_2(x,x') ={\cal K}(x-x').
\end{gather}

\subsection{Conductance}
\label{sec:Conc}
We have calculated the asymptotic solution of DMPK equation in the metallic 
regime and then obtained the universal terms of one- and two- point 
correlation functions $K_1(x)$ and $K_2(x,x')$ respectively for
all ten universality classes. The results are eq.\eqref{eq:A}, 
eq.\eqref{eq:B}, eq.\eqref{eq:C} and eq.\eqref{eq:D}.

Using these results, we can calculate the average and variance of 
arbitrary transport quantity $A$ (eq.~\eqref{eq:TQ});
\begin{itemize}
\item standard and B-dG classes
\begin{align}
\frac{\langle A \rangle_{sB}^0}{d}
&=\frac{\beta -2\kappa}{\beta}\left(\frac{f(0)}{4}+
\int_{0}^{\infty}{\rm{d}}x
\frac{f(x)}{4x^2+\pi^2}\right),\\
\frac{{\rm{Var}}A_{sB}}{d^2} &=\frac{1}{\beta\pi^2}
\int_0^{\infty}{\rm{d}}k
\frac{k|f(k)|^2}{1+\cot(\frac{1}{2}\pi k)},
\end{align}
where
\begin{gather}
f(k)=2\int_{0}^{\infty}{\rm{d}}xf(x)\cos kx.
\end{gather}
\item chiral class
\begin{align}
\frac{\langle A \rangle_{ch}^0}{d}
&=0,
\\
\frac{{\rm{Var}}A_{ch}}{d^2} &=\frac{2}{\beta\pi^2}
\int_0^{\infty}{\rm{d}}k
\frac{k}{1+\cot(\frac{1}{2}\pi k)}\left({f'}^2(k)+{f''}^2(k)\right),
\label{eq:fact}
\end{align}
where
\begin{align}
f'(k)&=\int_{-\infty}^{\infty}{\rm{d}}xf(x)\cos kx,\\
f''(k)&=\int_{-\infty}^{\infty}{\rm{d}}xf(x)\sin kx.
\end{align}
\end{itemize}
Into the above formulae we
substitute $f(x)$ corresponding to the conductance $G$,
the shot-noise power $P_{\rm{shot}}$ and the conductance for
the system having normal-superconducting junction 
$G_{\rm{NS}}$ which are given by eqs.~\eqref{eq:G} --~\eqref{eq:Gns}.
The universal values of the average and variance 
for each universality class are summarized in Table~\ref{tab:b}.

It is noted in Table~\ref{tab:b} that in standard and B-dG classes,
the universal values of the variances do not depend on $\kappa$, 
which characterizes the B-dG class. This is because in the asymptotic 
solution~\eqref{eq:AsB}, the constant $\kappa$ appears not in
the Jastrow type function
 $|\sinh^2x_i-\sinh^2x_j|$ 
but in one-body potential of $x_i \sinh2x_i$.  Considering this fact
with eq.~\eqref{eq:K2} and eq.~\eqref{eq:sBF}, we can easily find out
the independence of $\kappa$. 

We also notice that the variances for chiral class are twice as large
as those for standard and B-dG classes.   
This comes from the fact that $f(x)$'s in eqs.~\eqref{eq:G}--~\eqref{eq:Gns}
are all even function. Therefore, we have
\begin{gather}
\label{eq:}
\frac{{\rm{Var}}A_{\text{ch}}}{d^2}=2\frac{{\rm{Var}}A_{\text{sB}}}{d^2},
\end{gather}
due to $f(k)=f'(k)$ and $f''(k)=0$.

\vspace{7mm}

\begin{table}[htbp]
\begin{center}
\renewcommand{\arraystretch}{1.5}
 \begin{tabular}{|c|c|c|c||c|c|c|}         \hline
      & $\langle G\rangle$ &$\langle P\rangle$ &$\langle G_{NS}\rangle$ &var$G$&var$P$&
var$G_{NS}$\\ \hline\hline
standard and B-dG &$\frac{1}{3}\frac{\beta -2\kappa}{\beta}$&$\frac{1}{45}\frac{\beta -2\kappa}{\beta}$&$\left(2-\frac{8}{\pi^2}\right)\frac{\beta -2\kappa}{\beta}$ &$ \frac{2}{15}\frac{1}{\beta}$ &$ \frac{46}{2835}\frac{1}{\beta}$ &$\frac{64}{15}\left(1-\frac{45}{\pi^4}\right)\frac{1}{\beta}$ \\\hline
  chiral & $0$ &$0$ &$0$ &$\frac{4}{15}\frac{1}{\beta}$&$\frac{92}{2835}\frac{1}{\beta}$&$\frac{128}{15}\left(1-\frac{45}{\pi^4}\right)\frac{1}{\beta}$ \\ \hline
\end{tabular}

    \caption{The universal values of transport quantities.
We note here again that $G_{NS}$ can be expressed as eq.~\eqref{eq:Gns}
only if the time-reversal symmetry is not broken. Thus in the last column
these values are not correct for $\beta =2$ in standard and chiral classes
and C and D case in B-dG classes. See Table~\ref{tab:a} for the value of $\beta$
and $\kappa$.}
    \label{tab:b} 
  \end{center}
\end{table}

\section{Conclusion}
We summarize the universal transport properties of disordered 
quantum wire which have been shown in this paper. 
These results agree with the known
results by use of 
moment expansion method for conductance and shot-noise power~\cite{MBF1,ImHi}
and exact solutions of DMPK equations for some classes~\cite{MBF2,BFGM}.
Furthermore, we have obtained the universal terms of transport quantities 
for general transport
quantities, eq.~\eqref{eq:TQ} and for all ten universality classes.
Those results are new and have not been reported.

The following properties are interesting and should be observed in experiments:
\begin{enumerate}
\item The universal terms of averages of the transport quantities
vanish in chiral classes due to eq.~\eqref{eq:B}
while in standard and B-dG classes they 
depend on the parameters $\beta$ and $\kappa$ in the form of 
$\frac{\beta-2\kappa}{\beta}$. 
\item The variances of transport quantities are universal. In B-dG
classes, however, the dependence on $\kappa$ which characterizes
the B-dG classes does not appear in them. 
\item In chiral class, the variances are exactly twice as those in the standard class. 
\end{enumerate}
\begin{center}
\sffamily\bfseries{Acknowledgment}
\end{center}
One of the authors(T.I.) thanks Dr. K. Hikami  
for valuable suggestions and helpful comments.


\end{document}